\documentclass[prc,twocolumn,showpacs]{revtex4}
\usepackage{graphicx}
\usepackage{amsmath, amsthm, amssymb}
\usepackage{multirow}
\usepackage{subfigure}
\usepackage{bm}
\begin{document}

\title{An AC Stark Gradient Echo Memory in Cold Atoms}
\author{B. M. Sparkes, M. Hosseini, G. H\'{e}tet, P. K. Lam, and B. C. Buchler}
\affiliation{ARC Centre of Excellence for Quantum-Atom Optics, Department of Quantum Science, The Australian National University, Canberra, Australian Capital Territory 0200, Australia}
\begin{abstract}
The burgeoning fields of quantum computing and quantum key distribution have created a demand for a quantum memory. The gradient echo memory scheme is a quantum memory candidate for light storage that can boast efficiencies approaching unity, as well as the flexibility to work with either two or three level atoms. The key to this scheme is the frequency gradient that is placed across the memory. Currently the three level implementation uses a Zeeman gradient and warm atoms. In this paper we model a new gradient creation mechanism - the ac Stark effect - to provide an improvement in the flexibility of gradient creation and field switching times. We propose this scheme in concert with a move to cold atoms ($\simeq1$ mK). These temperatures would increase the storage times possible, and the small ensemble volumes would enable large ac Stark shifts with reasonable laser power. We find that memory bandwidths on the order of MHz can be produced with experimentally achievable laser powers and trapping volumes, with high precision in gradient creation and switching times on the order of nanoseconds possible. By looking at the different decoherence mechanisms present in this system we determine that coherence times on the order of 10s of milliseconds are possible, as are delay-bandwidth products of approximately 50 and efficiencies over 90\%.
\end{abstract}
\maketitle
\section{Introduction}
\label{sec:introduction}
The development of a quantum memory - a device that can store conjugate quantum variables - is currently being driven by the field of quantum information processing, a field encompassing quantum computing and quantum cryptography. An ideal quantum memory for these applications would be 100\% efficient, with long and controllable storage times, high storage bandwidths, and delay-bandwidth products, and faithful retrieval of the stored information.\\
Their speed and lack of interaction with the environment make photons an ideal carrier for quantum information, unfortunately these same properties make them difficult to store. In recent years, however, much progress has been made towards the development of an optical quantum memory with techniques such as electromagnetically induced transparency (EIT), where recall efficiencies of over 40\% in atomic ensembles \cite{e38} and storage times of over 1 second in solid state systems \cite{e15} have been achieved. Another quantum memory candidate is the atomic frequency comb (AFC) scheme, based on photon echoes, where the re-alignment of atomic dipoles is required for the stored light to be re-emitted. Storage of coherent states with a mean photon number of approximately one has been demonstrated with AFC \cite{afc1} and a maximum efficiency of 35\% has been achieved for a fixed storage time on the order of microseconds (determined by the bandwidth of the system) \cite{afc8}.\\
The gradient echo memory (GEM) technique is also a photon echo based coherent memory. The key ingredient for GEM is a frequency gradient imposed along the memory, as illustrated in Figure \ref{fig:ac_gem}(a). Not only does this gradient define the bandwidth of the system and cause re-emission of the pulse in the forwards direction by its reversal (Figure \ref{fig:ac_gem}(b)), it also allows for 100\% retrieval of the stored pulse if it is monotonic \cite{g2}. By altering the gradient, spectral manipulation of the pulse is possible \cite{g6}. In two level solid state systems an efficiency of 69\% for GEM has been demonstrated \cite{g8}. This work was carried out using an electric field (i.e. dc Stark) gradient created by placing four electrodes at the corners of the ensemble.\\
With a two level system the storage time of GEM will be limited by the decay rate from the excited state. By moving to a three level $\Lambda$ system in the far-detuned regime, as shown in Figure \ref{fig:ac_rb}(a), an effective two level atom is created. This is known as $\Lambda$-GEM, and apart from the increase in storage times achievable \cite{g1}, it also allows for pulse resequencing \cite{g5}. The decay rate of this system is now determined by the ground state decoherence rate.\\
Previous experimental work on $\Lambda$-GEM has used warm (65-70$^o$ C) gas cells, of length 7.5-20 cm and diameter 2.5 cm, containing $^{87}$Rb. This set-up has achieved efficiencies of up to 41\% \cite{g5} and coherence times of 20~$\mu$s. The frequency shift for the rubidium ensemble was created using a magnetic field (i.e. Zeeman) gradient, as alkali elements do not have a linear dc Stark shift. Using magnetic fields created by applying currents to solenoids wrapped around the gas cell, as in the above work, there is a lack of precision control over the gradient. Transient fields, which occur during gradient switching due to the inductance of the coils, also limit the switching time and can affect the rephasing process.\\
One option for improving the gradient creation and control would be to move away from magnetic fields, and the coils necessary to create them, to an ac Stark (acS) shift. This would allow for an all-optically controlled quantum memory. Another option for improvement is to move from warm to cold atoms. Due to the small decoherence rates in cold atomic ensembles this would allow for longer storage times and large on-resonance optical depths, due to the increase in density of the atoms. Implementing these improvements in concert would be beneficial as the acS effect is intensity dependent and cold atoms can be persuaded to occupy small volumes, reducing the laser power necessary.\\ 
This paper investigates the feasibility of using an acS generated frequency gradient for $\Lambda$-GEM in an ensemble of cold $^{87}$Rb atoms. After an overview of GEM theory in Section \ref{sec:ac_GEM_overview}, the main body of the paper (Section \ref{sec:ac_stark_shift}) discusses the proposed experimental implementation, including the theory behind the ac Stark shift and how it would apply to the creation of a frequency gradient across an ensemble of alkali atoms, as well as the optimal experimental parameters. Finally, factors that may limit this scheme will be discussed in Section \ref{sec:ac_limiting_factors} such as coherence times and maximum efficiencies possible.\\
\section{Gradient Echo Memory Theory}
\label{sec:ac_GEM_overview}

\begin{figure}[!ht]
\begin{center}
\includegraphics[width=\columnwidth]
{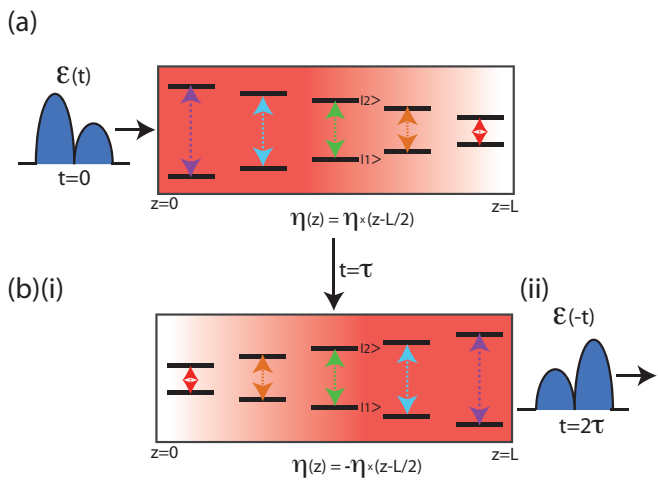}
\caption{The GEM Scheme. (a) At time $t=0$ a pulse with envelope $\mathcal{E}(t)$ enters the atomic ensemble of two level atoms with ground state $\left|1\right\rangle$ and excited state $\left|2\right\rangle$ and with a linear frequency gradient $\eta(z) = \eta \times (z-L/2)$ applied across them. (b)(i) At time $t=\tau$ the gradient is switched from $\eta \rightarrow -\eta$ causing a rephasing of the atomic dipoles and (ii) a release of the pulse, time reversed, occurring at time $t=2\tau$.}
\label{fig:ac_gem}
\end{center}
\end{figure}

\begin{figure}[!ht]
\begin{center}
\includegraphics[width=\columnwidth]
{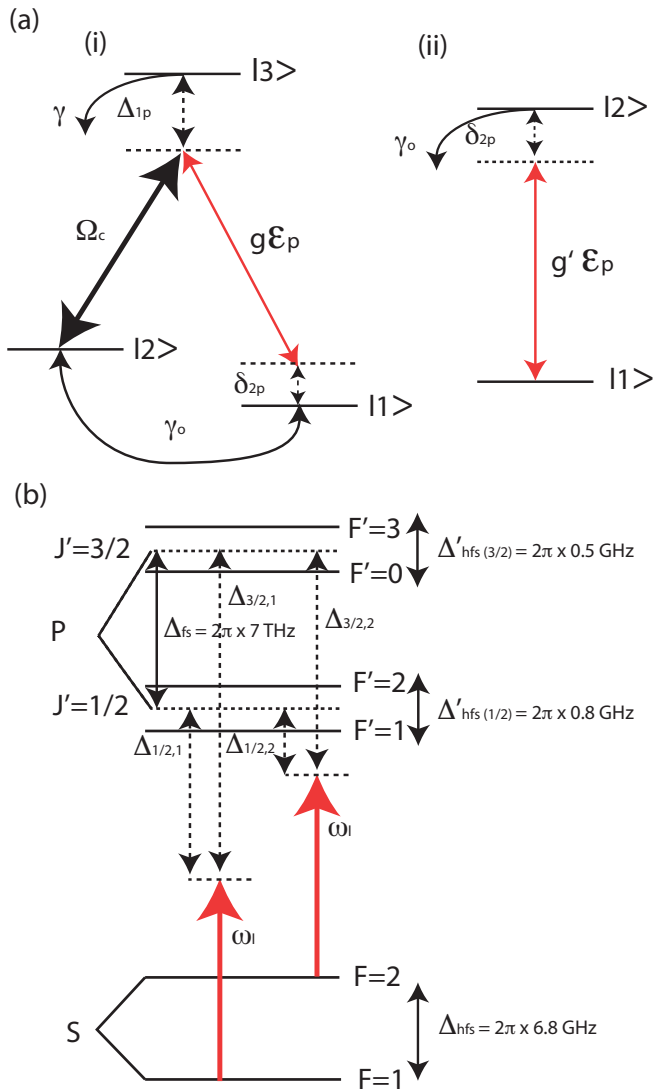}
\caption{Atomic Level Structures. (a)(i) The three-level system used for $\Lambda$-GEM showing the decay rate $\gamma$ from the excited state $\left|3\right\rangle$, the decoherence rate $\gamma_o$ between the two ground states $\left|1\right\rangle$ and $\left|2\right\rangle$, the one-photon detuning $\Delta_{1p}$, the two-photon detuning $\delta_{2p}$ and the coupling strength between the two levels $g$. In the presence of a strong coupling field with Rabi frequency $\Omega_c$ and weak probe field $\mathcal{E}_p$ this become equivalent to the two-level system shown in (ii), where the effective coupling strength $g' = g \Omega_c/\Delta_{1p}$. (b) The level structure of rubidium 87 showing the hyperfine splittings $\Delta_{hfs}$ between $F$ levels for both ground and excited states, as well as the fine-structure splitting $\Delta_{fs}$ between the two excited states. Also shown are the detunings $\Delta_{J',F}$ for a given laser frequency $\omega_{l}$ (see definitions in text).}
\label{fig:ac_rb}
\end{center}
\end{figure}

Consider a collection of $N$ two-level atoms with ground state $\left|1\right\rangle$ and excited state $\left|2\right\rangle$ with a resonant frequency $\omega_o$ and excited state decay rate $\gamma$, as shown in Figure \ref{fig:ac_rb}(a)(ii). We can then define atomic operators $\hat{\sigma}_{ij} = \left|i\right\rangle\left\langle j \right|$ and the atom-light coupling strength between the two levels
\begin{equation}
g=\sqrt{\frac{\omega_o}{2 \hbar \epsilon_o V}} \mu_{12},
\label{eq:ac_coupling_strength}
\end{equation}
where $\mu_{12}$ is the dipole transition strength between the two levels due to the presence of a light field with an interaction volume $V$.\\
If a linear frequency gradient is applied along an ensemble of atoms of length $L$, then the detuning from resonance of the atoms (i.e. the two-photon detuning $\delta_{2p}$ - see Figure 2(a)(ii)) will be given by $\delta_{2p}(z) = \eta \times (z-L/2)$. When a light field with a slowly varying envelope operator $\hat{\mathcal{E}}(z,t)$ and centre frequency $\omega_o$ is sent into such an ensemble, the equations that govern the storage of the light, in a moving frame at the speed of light ($t \rightarrow t-z/c$) and in the weak probe regime ($\left\langle \hat{\sigma}_{11}\right\rangle \approx 1$), are \cite{g2}
\begin{eqnarray}
\partial_t \hat{\sigma}_{12}(z,t) & = & -[\gamma/2 + i\eta (z-L/2)]\hat{\sigma}_{12}(z,t)\nonumber \\
& & + ig\mathcal{E}(z,t) \\
\partial_z \hat{\mathcal{E}}(z,t)& = & i\frac{g N}{c} \hat{\sigma}_{12}(z,t).
\label{eq:GEM_storage}
\end{eqnarray}
The efficiency of this writing stage of the memory is given by $\epsilon_{w} = 1-exp(-2\pi d')$ \cite{c9}, determined by the effective optical depth $d' = g^2 N/(c \eta)$. This assumes that the bandwidth of the pulse is smaller than the bandwidth of the ensemble given by $\mathcal{B}_s = \eta L$ and also that $\mathcal{B}_s \gg \gamma$.\\
To recall the pulse the gradient must be switched from $\eta \rightarrow - \eta$. This causes a rephasing of the dipoles, and therefore a time-reversal of the initial storage process, resulting in the emission of a photon echo from the memory in the forward direction. If, as shown in Figure \ref{fig:ac_gem}, the input pulse enters the memory at time $t=0$ and the field is switched at $t=\tau$, then the output pulse will be released at $t=2\tau$. Due to the reversal process the output pulse will be a mirror image of the input pulse with respect to time, i.e. $\hat{\mathcal{E}}(z,t) \rightarrow \hat{\mathcal{E}}(z,-t)$ as shown in Figure \ref{fig:ac_gem}(b)(ii). This can also be explained using the polariton description of the storage process presented in \cite{g3}.\\
If the gradient is monotonic along the memory then the read efficiency $\epsilon_{r} = \epsilon_{w}$ to give a total read/write efficiency for the memory of
\begin{equation}
\epsilon_{rw} = \left[1-exp(-2 \pi d')\right]^2,
\label{eq:ac_readwrite_efficiency}
\end{equation}
which will approach 100\% for large optical depths $d' \rightarrow 1$. If the gradient is not monotonic the pulse will be partially re-absorbed as it leaves the memory, lowering the recall efficiency, with a maximum efficiency of 54\% possible with no gradient \cite{c2}. Apart from the read-write efficiency there is also the storage efficiency $\epsilon_s$ to consider, which depends on the decay rate from the excited state and is of the form $\epsilon_s = exp(-\gamma t)$.\\

In previous work \cite{g1,g5} we have made use of the well known equivalence between a far detuned $\Lambda$ system, with a strong coupling field and a weak probe field, and a two level system. This equivalence is illustrated in Figure \ref{fig:ac_rb}(a), showing the coupling field Rabi frequency $\Omega_c$, probe field $\mathcal{E}_p$, and the detuning for both detuned from the excited state $\Delta_{1p}$ (the one-photon detuning). The advantage of the $\Lambda$ system (with $g$ now dependent on $\mu_{13}$ for the probe field and $\mu_{23}$ for the coupling field) over the two level one is that the decay rate from this excited state $\left| 2 \right\rangle$ is now limited by the decoherence rate $\gamma_o \ll \gamma$. For large on-resonance, unbroadened, optical depths ($d=g^2 N L/(c \gamma) \gg 1$) the conditions for this equivalence are \cite{e8}: (i) the system being far-detuned from resonance $\left|\Delta_{1p}\right| \gg d\gamma$; and (ii) $T \gamma d \gg 1$, where $T$ is the fastest timescale of the system, usually either dependent on the pulse length $t_p$ or $\Omega_c$. The equations of motion for the $\Lambda$ system then become equivalent to the two level equations of motion except for $g \rightarrow g' = g\Omega_c/\Delta_{1p}$. The effective optical depth for the equivalent two level system will therefore be
\begin{equation}
d' = \frac{g^2 N}{c \eta}\left(\frac{\Omega_c}{\Delta_{1p}}\right)^2.
\label{eq:ac_optical_depth}
\end{equation}
\section{AC Stark Shift Proposal}
\label{sec:ac_stark_shift}

\begin{figure}[!ht]
\begin{center}
\includegraphics[width=\columnwidth]
{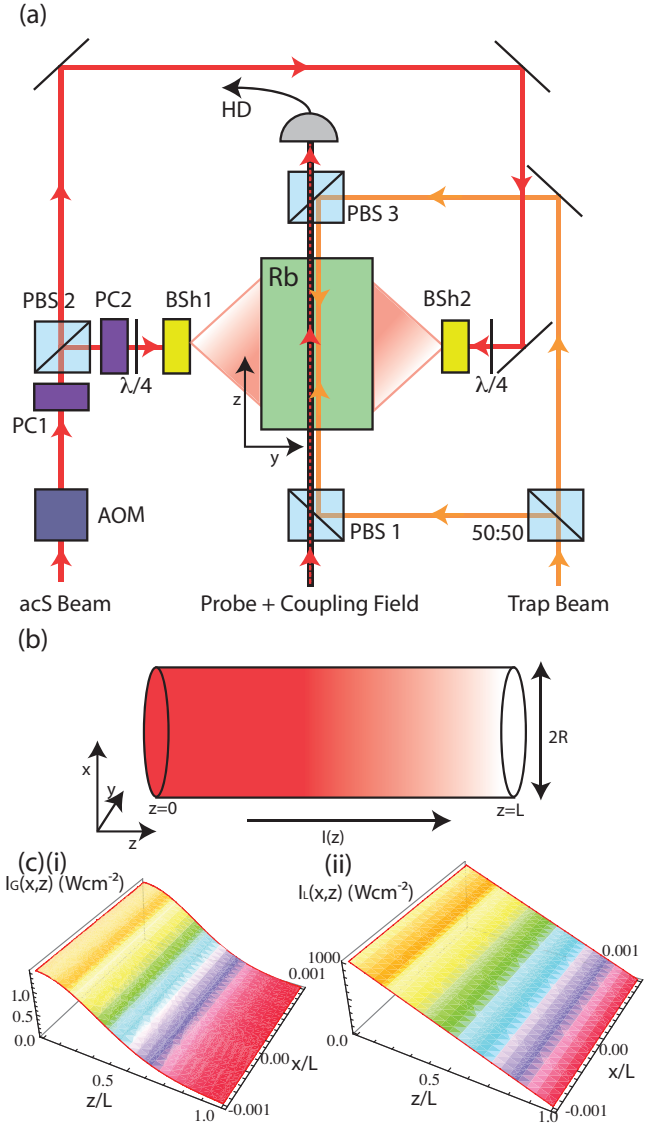} 
\caption{The Experiment. (a) Envisioned set-up for $\Lambda$-GEM experiment using cold atoms and an ac Stark gradient, with: Rb - atomic ensemble of rubidium 87; 50:50 - 50:50 beam splitter; PBS - polarizing beam splitter; PC - Pockels cell; BSh - Beam shaper; and HD - heterodyne detector. The polarizations of the acS and trapping fields are assumed to be linear, while optimal probe and coupling field polarizations will be discussed in Section \ref{sec:ac_switching_protocols}. (b) Side-on view of the cylindrical atomic ensemble showing the direction of the intensity gradient $I(z)$ and radius $R$. (c)(i) Gaussian and (ii) linear intensity profiles for the acS field per unit power over the ensemble. Here $L$ has been taken to be 1 cm and $R = 10$ $\mu$m, determined by the radius of the trapping laser.}
\label{fig:ac_setup}
\end{center}
\end{figure}

The envisaged experiment is shown in Figure \ref{fig:ac_setup}(a). The probe and coupling fields would be created similarly to the method described in \cite{g5}, with the coupling field being 6.8 GHz detuned from the probe to allow both $S_{1/2}$ ground states of $^{87}$Rb to be used (see Section \ref{sec:ac_rubidium_structure}).\\
There are three main components to this set-up: (i) cold atom storage; (ii) ac Stark gradient creation and (iii) switching (including probe and coupling field orientations). These will each be discussed in turn after firstly looking at the structure of rubidium, which will be used as the storage medium, and the ac Stark theory.\\
\subsection{Rubidium Structure}
\label{sec:ac_rubidium_structure}
Rubidium was used for previous experimental work as it provides a convenient working wavelength for the memory. Rubidium 87, as with all alkali atoms, has a well known structure with, in this case, two hyperfine ground states $S_{1/2}, F=1,2$ and both a $D1$ and $D2$ transition, which correspond to P$_{1/2}$ and P$_{3/2}$ levels respectively with wavelengths of 795 and 780 nm \cite{Steck}. This structure is shown in Figure \ref{fig:ac_rb}(b), containing the hyperfine structure splittings $\Delta_{hfs}$ and $\Delta_{hfs}'(J')$ for the ground and excited states respectively, as well as the fine structure splitting $\Delta_{hf}$ between the two excited states.\\
The decay rate for the excited states $J'$ of $^{87}$Rb is $\gamma/2\pi \approx 6$ MHz, giving an excited state lifetime of approximately 30 ns. Also, the hyperfine splitting $\Delta_{hfs}/2\pi = 6.8$ GHz for the ground states and $\Delta_{hfs}'(3/2)/2\pi = 500$ MHz, $\Delta_{hfs}'(1/2)/2\pi = 800$ MHz for the excited states, with fine structure splitting $\Delta_{fs}/2\pi = 7$ THz.
\subsection{AC Stark Shift Theory}
\label{sec:ac_stark_shift_theory}
When light of intensity $I(r,z)$, frequency $\omega_l$ and polarisation $q=0,\pm1$ (corresponding to linear, right ($+$) and left ($-$) circular polarizations respectively) is shone onto an atom there will be a change in energy of the internal states. This is known as the ac Stark effect. For alkali atoms, with structure as shown in Figure \ref{fig:ac_rb}(b), this effect can be calculated for a given ground state $\left| g_i \right\rangle = \left|1/2,F,m_F\right\rangle$ from second order time-dependent perturbation theory to be \cite{Sakurai}
\begin{equation}
U_{F,m_F}(\omega_l,q,I) = \frac{I(r,z)}{2 c \epsilon_o \hbar} \sum_{a}\frac{\left|\left\langle a |e \hat{\mathbf{r}} \cdot \bm{\epsilon}_q|g_i\right\rangle\right|^2}{\omega_l - \omega_{ag_i}},
\label{eq:ac_energy_dip}
\end{equation}
where the sum is over all excited states $\left| a\right\rangle = \left| J', F',m_F' \right\rangle$, $\omega_{ag_i}$ is the frequency of the transition between $\left| a\right\rangle$ and $\left| g_i \right\rangle$ allowing us to define the detuning as $\Delta = \omega_l - \omega_{ag_i}$. This formula uses the rotating wave approximation, which is valid for detunings much smaller than the frequency of the transition, i.e. $\omega_l - \omega_{ag_i} \ll \omega_l + \omega_{ag_i}$. For simplicity we can write the above equation as $U_{F,m_F}(\Delta,q,I)=\bar{U}_{F,m_F}(\Delta,q)I(r,z)$, where $\bar{U}_{F,m_F}$ is the change in energy per unit intensity. An approximation to Equation \ref{eq:ac_energy_dip}, in the limit of $\Delta \gg \Delta_{hfs}'$, is given by \cite{rb11}
\begin{eqnarray}
U_{F,m_F}(\Delta,q,I) & \simeq & \frac{\pi c^2 \gamma I(r,z)}{2 \omega_o^3}\times \nonumber \\ 
& & \left( \frac{2 + qg_Fm_F}{\Delta_{3/2,F}} + \frac{1 - q g_F m_F}{\Delta_{1/2,F}} \right),
\label{eq:ac_energy_dip_approximation}
\end{eqnarray}
where $\gamma$ and $\omega_o$ are the averaged values of the two excited levels, $g_F$ is the Lande factor ($g_1 = -0.5$, $g_2 = -g_1$) and $\Delta_{J',F}$ is the detuning from the $S_{1/2}$ level ($F=1,2$) to either $J'=1/2,3/2$ levels, as illustrated in Figure \ref{fig:ac_rb}(b).\\
Apart from changing the energies of the atomic levels, there is the possibility that the atoms will absorb and then re-emit the light. This scattering of light by the atoms will affect not only the coherence time $\tau_{coh}$ achievable with the ensemble, but also the lifetime $\tau_{trap}$ of any trap that is used to contain them (see Section \ref{sec:ac_atom_trapping}). The scattering rate for a given ground state will be determined by \cite{Loudon}
\begin{eqnarray}
\Gamma_{F,m_F} (\omega_l,q,I)& = & \frac{I(r,z)}{6 \pi \epsilon_o^2 \hbar^3 c^4} \sum_{g_f}(\omega_l - \omega_{fi})^3 \times \nonumber \\
& & \left| \sum_{a,q_{sc}}\frac{\left\langle g_f| e \hat{\mathbf{r}} \cdot \bm{\epsilon}_{q_{sc}}|a\right\rangle \left\langle a| e \hat{\mathbf{r}} \cdot \bm{\epsilon}_{q}|g_i\right\rangle}{\omega_{a g_i} - \omega_l}\right|^2
\label{eq:ac_scattering_rate_full}
\end{eqnarray}
again in the rotating wave approximation, where $\left|g_f\right\rangle$ is the final state (which cannot be higher in energy than $E_{g_i}+\hbar \omega_l$ by conservation of energy), $\omega_{fi}$ is the frequency of the transition from state $\left|g_i\right\rangle$ to $\left|g_f\right\rangle$ (negative if $\left|g_f\right\rangle$ is lower in energy than $\left|g_i\right\rangle$), $q_{sc}$ is the polarisation of the scattered photon, and $\omega_{ag_i}$ is the frequency of the transition between states $\left|g_i\right\rangle$ and $\left|a\right\rangle$.\\
The above equation can be further simplified for alkali atoms if the detuning is much greater than the excited state hyperfine splitting $\Delta'_{hfs}$ to give
\begin{eqnarray}
\Gamma_{F,m_F} (\omega_l,q,I)&=& \frac{I(r,z)}{6 \pi \epsilon_o^2 \hbar^3 c^4} \sum_{g_f}(\omega_l-\omega_{fi})^3 \times \nonumber \\ 
& & \left|\frac{A_{1/2,g_i}}{\Delta_{1/2,F}} + \frac{A_{3/2,g_i}}{\Delta_{3/2,F}}\right|^2 \nonumber \\
&=& \bar{\Gamma}_{F,m_F}(\Delta,q) I(r,z),
\label{eq:ac_scattering_rate_simplified}
\end{eqnarray}
where $\bar{\Gamma}_{F,m_F}$ is the scattering rate per unit intensity, and 
\begin{equation}
A_{J',g_i} \equiv \sum_{a,q_{sc}}\left\langle g_f| e \hat{\mathbf{r}} \cdot \bm{\epsilon}_{q_{sc}}|a\right\rangle \left\langle a| e \hat{\mathbf{r}} \cdot \bm{\epsilon}_{q}|g_i\right\rangle
\label{eq:ac_definition}
\end{equation}
for all states $\left| a \right\rangle$ within the level $J'$. As can be seen from the above equations $\bar{\Gamma}_{F,m_F} \propto 1/\Delta^2$ while $\bar{U}_{F,m_F} \propto 1/\Delta$ and therefore, in the context of laser trapping, increasing the detuning for a constant trap depth will increase both $\tau_{coh}$ and $\tau_{trap}$.\\

\begin{figure}[!ht]
\begin{center}
\includegraphics[width=\columnwidth]
{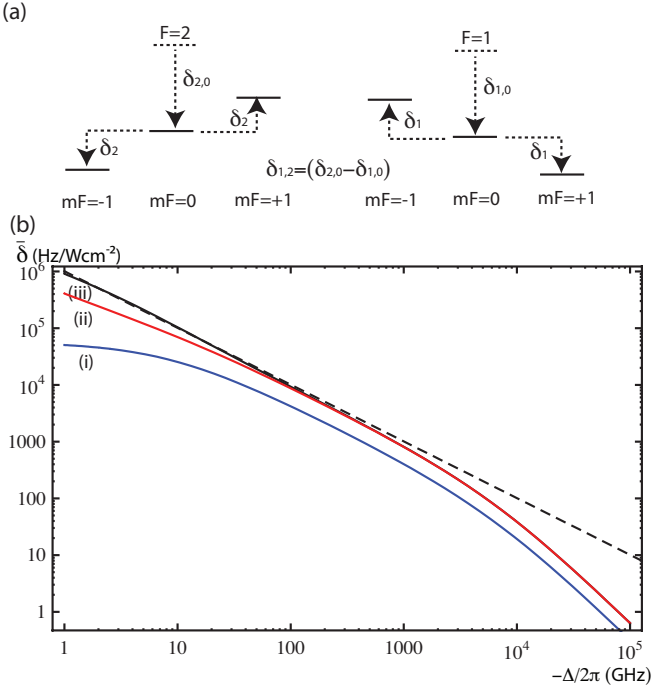} 
\caption{Frequency Splitting. (a) Illustration of $\delta_F$ splitting for the $m_F=\pm1, 0$ states of the $F=1,2$ levels. This is assuming a polarisation of $q=1$ for the acS field. $\delta_{1,2}$ is the difference between the splittings of the two $m_F=0$ states of the respective levels from their original positions. (b) Log-log plot of (i) $\bar{\delta}_1$, (ii) $\bar{\delta}_2+\bar{\delta}_1$, and (iii) $\bar{\delta}_t =\bar{\delta}_2+\bar{\delta}_1 + \bar{\delta}_{1,2}$ as a function of detuning $\Delta_{1/2,2}/2\pi$. The dashed line shows a fit to $\bar{\delta}_t$ for small detunings where $\bar{\delta}_t \propto 1/\Delta$. These traces were calculated using Equation \ref{eq:ac_energy_dip}.}
\label{fig:ac_mFsplitting}
\end{center}
\end{figure}

Here we are interested in the energy splitting (and therefore frequency splitting $h \Delta \nu =\Delta U$) along an ensemble of atoms. This will determine the bandwidth of the system $\mathcal{B}_s$. The splitting per unit intensity between two adjacent $m_F$ states in the same $F$ level can be defined to be
\begin{equation}
\bar{\delta}_F(\Delta,|q|) \equiv 1/h \left| \bar{U}_{F,0}(\Delta,q) - \bar{U}_{F,1}(\Delta,q)\right|,
\label{eq:ac_deltaf}
\end{equation}
with the total magnitude of the splitting given by $\delta_F(\Delta,|q|,I) = \bar{\delta}_F(\Delta,|q|)I(r,z)$. Combining the above equation with Equation \ref{eq:ac_energy_dip_approximation} we obtain
\begin{equation}
\bar{\delta}_F(\Delta,|q|) = \frac{ \pi c^2 \gamma}{2 \omega_o^3 h} \left| \frac{q g_F}{\Delta_{1/2,F}}\right|\left(1-\frac{\Delta_{1/2,F}}{\Delta_{3/2,F}}\right).
\label{eq:ac_deltaf2}
\end{equation}
The frequency splitting between the $m_F=0$ states of the $F=1$ and $2$ levels can similarly be found to be
\begin{eqnarray}
\bar{\delta}_{1,2}(\Delta) &\equiv & \frac{1}{h} \left( \bar{U}_{1,0}(\Delta) - \bar{U}_{2,0}(\Delta)\right) \nonumber \\
&= &\frac{ \pi c^2 \gamma}{2 \omega_o^3 h}\left(\frac{2(\Delta_{3/2,1}-\Delta_{3/2,2})}{\Delta_{3/2,2}\Delta_{3/2,1}} + \frac{\Delta_{1/2,1}-\Delta_{1/2,2}}{\Delta_{1/2,2}\Delta_{1/2,1}}\right) \nonumber \\
& = & \frac{\pi c^2 \gamma\Delta_{hfs}}{2 \omega_o^3 h}\left(\frac{2}{\Delta_{3/2,2}\Delta_{3/2,1}} + \frac{1}{\Delta_{1/2,2}\Delta_{1/2,1}}\right).
\label{eq:ac_delta12}
\end{eqnarray}
These frequency splittings are illustrated in Figure \ref{fig:ac_mFsplitting}(a) with the polarisation taken to be $q=1$ (i.e. right circularly polarized) as, if linear polarisation were to be used, $\bar{\delta}_{F} = 0$.\\
If, as was the case in \cite{g5}, one $m_F$ state in each of the hyperfine levels is used for the ground states $\left|1\right\rangle$ and $\left|2\right\rangle$ from Figure \ref{fig:ac_rb}(a), then we can define the total splitting in terms of $\bar{\delta}_{F}$ and $\bar{\delta}_{1,2}$ to be
\begin{equation}
\bar{\delta}_t(\Delta,q) \equiv \bar{\delta}_{1,2}(\Delta) -q\left(m_2\bar{\delta}_2(\Delta,|q|) + m_1\bar{\delta}_1(\Delta,|q|)\right).
\label{eq:ac_deltabar}
\end{equation}
Figure \ref{fig:ac_mFsplitting}(b) shows the absolute value of $\delta_t$ for $m_2 = -m_1 = -1$ and the relative contributions from the three terms above as a function of detuning for $q=1$. As can be seen, at large detunings $\Delta_{1/2,F} \gg \Delta_{hfs}$ the splitting between $m_F$ levels becomes approximately equal (i.e. $\bar{\delta}_1 \sim \bar{\delta}_2$) and the relative contribution of $\bar{\delta}_{1,2} \rightarrow 0$. This behaviour can be explained from Equations \ref{eq:ac_deltaf2} and \ref{eq:ac_delta12} above as $\Delta_{J',1} \simeq \Delta_{J',2}$ for $\Delta_{1/2,F} \gg \Delta_{hfs}$. \\
From Equation \ref{eq:ac_deltabar} we can calculate the field gradient to be $\eta(z) \simeq 2\pi \bar{\delta}_t \partial_z I(r,z)$ assuming negligible change for the intensity in the transverse ($r$) direction. Using this equation for $\eta$ allows the system bandwidth to be expressed in terms of $\bar{\delta}_t$ and $I$ as follows
\begin{eqnarray}
\mathcal{B}_s(\Delta,|q|I) &= &\frac{1}{2\pi}\int_{0}^{L}|\eta(\Delta,q,I)| dz \nonumber \\
& =&\left|\bar{\delta}_t(\Delta,q) \left(I(z=L)-I(z=0)\right)\right|.
\label{eq:ac_system_bandwidth}
\end{eqnarray}
\subsection{Atom Trapping}
\label{sec:ac_atom_trapping}
One of the most common methods of cooling atoms to millikelvin temperatures is a Magneto-Optical Trap (MOT). It would seem an obvious suggestion to also use the MOT to confine the atoms during the memory process. However, a MOT has 6 circularly polarized beams detuned a few $\gamma$ below resonance and a magnetic field gradient on the order of 1 G/cm \cite{mot6,Foot}. The scattering rates associated with these beams and the $m_F$ splitting that would occur due to the magnetic field rule out the use of a MOT when storing information in an ensemble of atoms for $\Lambda$-GEM. If the MOT is turned off, however, the atoms are free to diffuse away from the interaction area and, though storage of light with a coherence time of 1 ms has been achieved using this method \cite{oqm14}, if longer storage times are to be achieved another form of trapping must be used.\\
One way to achieve this trapping is to use a far off resonance, red-detuned laser to create a dipole trap. This trapping field must be $\pi$ polarized and detuned on the order of $\Delta_{fs}$ or higher so as not to affect the different $m_F$ splitting, and therefore the gradient, along or across the ensemble. Achieving $\pi$ polarisation requires the addition of a constant dc magnetic field to create a quantisation axis. Such a beam, tightly focused and collimated, should provide sufficient trapping in the radial direction but much weaker trapping along the ensemble, as $F_{t} = -\nabla (U_t) = -\bar{U}_t \nabla (I(r,z))$ and $ \partial_r I \gg \partial_z I$. The radius of the ensemble will then become approximately equal to the waist of the trapping laser $w_{t}$, which can be reduced down to the order of 10 $\mu$m. This is desirable for our system as a reduction in area for the acS gradient creating laser ($\mathcal{A}_{acS} \geq 2 L w_t$) will lead to an increase in the intensity, and therefore bandwidth, for a given power (see Section \ref{sec:ac_beam_shaping}). Realistic experimental parameters give $L=1$ cm \cite{mot1,e29}.
To obtain the longest storage times possible the dipole laser should be detuned far from resonance. Assuming an easily obtainable experimental wavelength of $\lambda_{t} = 1064$ nm and a waist of $w_t = 10$ $\mu$m, a trap of depth $U_t = 1$ mK is achievable with a power of 1.5 W, giving a maximum scattering rate of $\Gamma_{t} = 4$ s$^{-1}$ \cite{rwa}.\\
To increase the trapping in the $z$ direction, one solution is to split the trapping field, using a 50:50 beam splitter, and send it into the ensemble from both directions to create a standing wave trap, with trapping maxima occurring at $\lambda_{t}/2$. The difference in detuning between two sites being 
\begin{equation}
\Delta \delta_t = \frac{2\pi \mathcal{B}_s \lambda_{t}}{2L}
\label{eq:ac_deltadelta}
\end{equation}
assuming a linear intensity profile.\\
Dipole traps have already been used for pulse storage, with storage times up to a few milliseconds achieved \cite{oqm15}. However, coherence times up to seconds are theoretically possible \cite{mot43}.\\
\subsection{Gradient Creation}
\label{sec:ac_gradient_creatin}
\subsubsection{Wavelength Selection}
\label{sec:ac_wavelength}

\begin{figure}[!ht]
\begin{center}
\includegraphics[width=\columnwidth]
{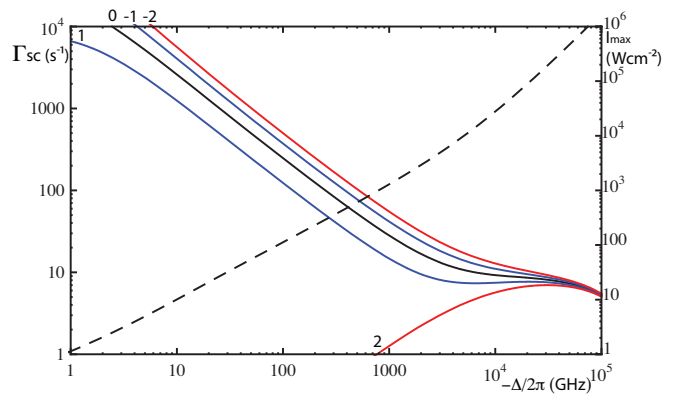}
\caption{The scattering rate required to produce a 1 MHz frequency splitting, for all $m_F$ states of the $F=2$ level, as a function of detuning $\Delta_{1/2,2}/2\pi$ (solid lines, left scale). Also shown is the intensity of the acS field required to produce the 1 MHz splitting (dashed line, right scale). These were calculated using Equations \ref{eq:ac_energy_dip} and \ref{eq:ac_scattering_rate_full} without using the rotating wave approximation.}
\label{fig:ac_acbeamscatter}
\end{center}
\end{figure}

A critical parameter that must be determined is the wavelength of the acS laser to be used, as this will set a limit on the maximum frequency splitting possible for a given laser power and intensity distribution, as well as the scattering rate of the system. To optimize the wavelength we must balance the desired behaviour (i.e. frequency splitting) with the undesired effect of light scattering by the atoms determined by $\bar{\Gamma}_{F,m_F}$. It will be assumed that the acS field will be circularly polarized to maximize $\bar{\delta}_t$ for large detunings (see Section \ref{sec:ac_stark_shift_theory}).\\
From Equation \ref{eq:ac_deltaf2} we can see that the splitting, and therefore bandwidth, depends not only on $1/\Delta$ but also on the ratio of $\Delta_{1/2,F}/\Delta_{3/2,F}$. As the detuning becomes large compared to $\Delta_{fs}$, $\Delta_{1/2,F}/\Delta_{3/2,F} \rightarrow 1$ as $\Delta_{3/2,F} = \Delta_{1/2,F} + \Delta_{fs}$, and therefore the bracketed term in Equation \ref{eq:ac_deltaf2} heads to zero. This means that $\bar{\delta}_F$ will head towards zero faster than $1/\Delta$ for large detunings. This can be seen from the difference in Figure \ref{fig:ac_mFsplitting}(b) between $\bar{\delta}_t$ and the dashed line, which shows the path $\bar{\delta}_t$ would take if it had only a $1/\Delta$ dependence. This deviation occurs on the order of THz detuning and is important as it leads to Figure \ref{fig:ac_acbeamscatter}. This shows the scattering rates of different $m_F$ states of the $F=2$ level for a bandwidth of 1 MHz, assuming the same splitting arrangement used for Figure \ref{fig:ac_mFsplitting}(b) and $q=1$, i.e. 
\begin{equation}
\Gamma_{2,m_F}(\Delta ,1) = \frac{10^6 \bar{\Gamma}_{2,m_F}(\Delta ,1)}{\bar{\delta}_t(\Delta,1)},
\label{eq:ac_optimal_scatering}
\end{equation}
with the corresponding intensity profile also shown \cite{rwa}.\\
From this figure it can be seen that the behaviour for the $m_F = 0,-1,-2$ levels is similar to that described for laser trapping in Section \ref{sec:ac_stark_shift_theory}, i.e. decreasing scattering rate with increasing detuning for a set trap depth. However, for $m_F = 1$ a minimum for the scattering rate is present at approximately 5 THz, with the scattering rate flattening out until 20 THz when it starts to decrease again. The $m_F = 2$ state shows even more peculiar behaviour, with a maximum appearing at approximately 20 THz. A positive slope indicates that $\bar{\delta}_t$ is decreasing faster than $\bar{\Gamma}$. This is to do in part with the dependence of $\bar{\delta}_t$ on $\Delta_{1/2,F}/\Delta_{3/2,F}$ discussed above, as well as the relative strengths of $A_{3/2,g_i}$ and $A_{1/2,g_i}$ and the levels which are allowed to contribute to $\bar{\Gamma}$. All scattering rates converge at detunings much larger than $\Delta_{fs}$. \\
The scattering rate for the $F=1$ states are approximately the same in magnitude but for the opposite $m_F$ state, with the $m_F = -1$ ground state containing this minima. In both cases the unusual behaviour occurs for the states which are raised in energy with respect to $m_F=0$.\\
Due to the small probe approximation mentioned in Section \ref{sec:ac_GEM_overview}, most of the population will remain in the $F=1$ state during the storage and retrieval processes. It would therefore be advantageous to make use of the minima and the $F=1, m_F=-1$ state as, though lower scattering rates can be achieved at much larger detunings, the laser intensities required become impractical. For instance, to achieve the same, or smaller, scattering rate than at the minima (approximately 11 s$^{-1}$) requires a detuning greater than 40 THz. At this detuning, it requires nearly 10 times the laser intensity to achieve the same bandwidth as at the minima. For this minima detuning $\bar{\delta}_1 \simeq \bar{\delta}_2 \simeq 50$ Hz/Wcm$^{-2}$ and $\bar{\delta}_{1,2} \ll \bar{\delta}_F$.
\subsubsection{Beam Shaping}
\label{sec:ac_beam_shaping}
The optimal intensity distribution - that which gives the maximum bandwidth for a given power $P_o$ - depends firstly on the orientation of the ensemble. As discussed in Section \ref{sec:ac_atom_trapping}, the ensemble will be cylindrical with $L=1$ cm and $R=w_t = 10$ $\mu$m. \\
One of the simplest intensity profiles to produce is a focused Gaussian, where 
\begin{equation}
I_G(r',y) = I_o exp\left[ -2 r'^2/w_o^2\right],
\label{eq:ac_gaussian_intensity}
\end{equation}
with $w_o$ the waist of the beam occurring at position $y=0$ (the centre of the ensemble, assuming the acS field is propagating along the $y$ axis), $I_o$ is the maximum intensity and $r'$ is the radial component in the $x$-$z$ plane. The only constraint on the acS beam shape is that it is monotonic along the ensemble (see Section \ref{sec:ac_GEM_overview}). In the case of a Gaussian beam, this means that over half the power will be lost, limiting the total splitting possible for a given laser power, with $I_o = 2 P_o /(\pi w_o^2)$.\\
To use the largest fraction of the beam possible, we can set $w_o = 2L/3$ so that if the centre of the beam is at $z=0$, approximately 99\% of the remaining intensity will fall on the ensemble. Figure \ref{fig:ac_setup}(c)(i) shows this intensity profile along and across the ensemble for unit power. As the slope $\eta$ is not linear, different frequency components will be stored with different efficiencies, as can be seen from Equation \ref{eq:ac_optical_depth}. In the case described above, where $R \ll L$, the frequency change along the $x$ direction for a set $z$ position will be negligible.\\
If one of the experimentally simplest intensity profiles for the acS beam is the one described above, then one of the most efficient will be one that covers only the ensemble, decreasing linearly from maximum intensity $I_o$ at one end to 0 at the other and with no change in intensity in the $x$ direction. Such an intensity profile will be of the form
\begin{eqnarray}
I_L(r',y) & = & I_o \left(1-z/L\right), \left|x\right| \leq R \nonumber \\
& = & 0, \left|x\right| > R
\label{eq:ac_linear_intensity}
\end{eqnarray}
over the length of the ensemble $0\leq z \leq L$. In this case $I_o = P_o/(LR)$ and therefore the maximum intensity, and bandwidth, achievable will be over 500 times larger than a Gaussian beam with the same power due to the smaller area the beam occupies. This can be seen from Figure \ref{fig:ac_setup}(c)(ii), which shows the intensity profile $I_L$ for unit power along and across the ensemble.\\
To change an initially Gaussian beam to one with an intensity profile like that from Equation \ref{eq:ac_linear_intensity} requires a beam shaper. These devices (for instance deformable mirrors, phase plates or liquid crystal spatial light modulators - LCSLMs) can be highly efficient ($\epsilon > 0.9$) and can be used to create nearly any desired beam shape with resolution on the order of 1000$\times$1000 pixels for LCSLMs. This not only provides us with a method for optimizing the acS laser intensity profile, but would also allow for spectral manipulation of the pulse to be carried out with the ability to produce complex gradients and switching arrangements such as those described in \cite{g6}.\\
\subsection{Switching Protocols}
\label{sec:ac_switching_protocols}
There are two components that make up the switching protocol: which $m_F$ states within the two $F$ levels are to be used; and the method for switching the acS field. These will be discussed in turn.
\subsubsection{Selecting $m_F$ Levels}
\label{sec:ac_selecting_mf_levels}
Selection rules determine which $m_F$ states of the two different levels can be used, depending on the polarizations of the probe and coupling fields ($q_p$ and $q_c$ respectively), with the total change in angular momentum being given by 
\begin{equation}
\Delta m_F = m_2 - m_1 = q_p-q_c.
\label{eq:ac_angular_momentum_change}
\end{equation}
As $m_1 = -1$ has already been decided upon as state $\left| 1\right\rangle$ (see Section \ref{sec:ac_wavelength}), the above equation can be re-arranged to give
\begin{equation}
m_2 = -1 + q_p - q_c.
\label{eq:ac_m2}
\end{equation}
We can determine the probe and coupling polarizations that will produce this maximum splitting by substituting the above relationship into Equation \ref{eq:ac_deltabar} to give
\begin{eqnarray}
\bar{\delta}_t & = & \bar{\delta}_{1,2} + \nonumber \\
& & q\left(\bar{\delta}_1 - (-1 + q_p - q_c)\bar{\delta}_2\right) \nonumber \\
&\simeq & q\bar{\delta}_F\left(2+ q_c - q_p\right),
\label{eq:ac_deltat_approx}
\end{eqnarray}
with the approximations $\bar{\delta}_1 \simeq \bar{\delta}_2 \gg \bar{\delta}_{1,2}$ and $|q| \neq 0$ discussed previously. This reveals that $\bar{\delta}_t$ can range from 0 to $3 \bar{\delta}_F$ (remembering $|m_2| \leq 2$) with the latter being possible for either $q_c = 1, q_p = 0$ or $q_c = 0, q_p = -1$.\\
Using the previous value of $\bar{\delta}_F$ determined in Section \ref{sec:ac_wavelength}, this would give a system bandwidth of $\mathcal{B}_s = 150$ Hz/Wcm$^{-2}$. If we combine this value with the maximum intensity achievable with the intensity profile $I_L$ we find $\mathcal{B}_s = 150$ kHz/W and therefore system bandwidths on the order of 1 MHz would be obtainable with less than 10 W of acS laser power using the optimal detuning. Also, using one of the above optimal level schemes with $\delta_t = 3 \delta_F$, the scattering rate per 1 MHz splitting given in Equation \ref{eq:ac_optimal_scatering}, which was calculated using $\delta_t = 2\delta_F$, will be reduced as the intensity required to reach the same bandwidth will be reduced by a factor of $2/3$.\\
The polarizations for the probe and coupling fields mean that they cannot be completely separated on a polarizing beam splitter. To detect only the probe at the end of the memory a frequency selective measurement, such as heterodyne detection, could be used.\\

\begin{figure}[!ht]
\begin{center}
\includegraphics[width=\columnwidth] 
{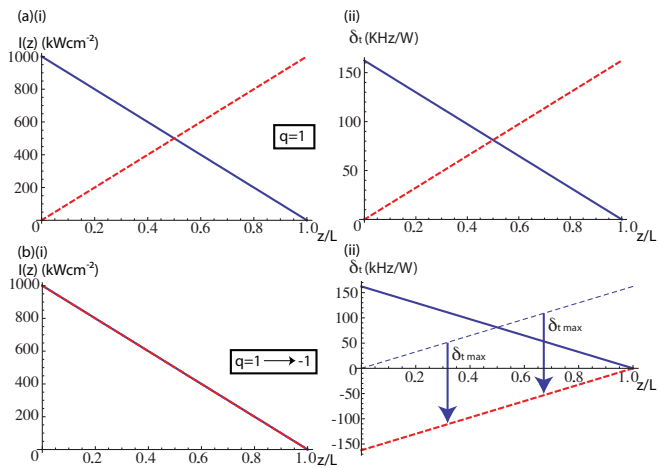} %
\caption{Switching Methods. (a)(i) Intensity profile per unit power and (ii) corresponding frequency splitting along the ensemble for switching method 1, with initial (solid line) and final (dashed line) gradients shown. Here the polarisation of the acS field $q=1$ for both initial and final gradients. (b)(i) Intensity profile per unit power and (ii) corresponding frequency splitting for switching method 2, with initial (solid line) and final (dashed line) gradients shown. Here the initial polarisation of the acS field is $q=1$, while the final polarisation is $q=-1$. This causes an frequency offset to the gradient of $-\delta_{t,max}$. Both (a) and (b) use the linear intensity profile $I_L$ (see Equation \ref{eq:ac_linear_intensity}) and the maximum $\bar{\delta}_t \simeq 3 \bar{\delta}_F$.}
\label{fig:ac_states}
\end{center}
\end{figure}

\subsubsection{Field Switching Method}
\label{sec:ac_field_switching_method}
To cause rephasing of the atomic dipoles we must be able to invert the detunings of the atoms. If we make the same assumption as in previous sections, i.e. $|q| = 1$, two switching methods become readily apparent.\\
The first method involves reversing the intensity profile along the ensemble $I(z) \rightarrow I(L-z)$. This is equivalent to the field switching method used in \cite{g2,g1,g5} in that by reversing the intensity profile about the centre of the trap ($z=L/2$) the detunings are also reversed about this point, i.e. $\delta_t(z) \rightarrow \delta_t(L-z)$. The intensity profiles for this method of switching are shown in Figure \ref{fig:ac_states}(a)(i), with the corresponding frequency gradients shown in Figure \ref{fig:ac_states}(a)(ii). This process involves no change in frequency of the stored pulse with respect to the input pulse.\\
The second method involves switching the polarisation of the field $q \rightarrow -q$ while keeping the same intensity gradient, as shown in Figure \ref{fig:ac_states}(b)(i). This is a slightly more complicated process as the detunings are no longer reversed around the centre of the ensemble, with $\delta_t(z) \rightarrow -\delta_t(z)$. This method still results in an echo being produced. However, the stored pulse will now be frequency shifted with respect to the input pulse, as can be seen from the corresponding frequency gradients shown in Figure \ref{fig:ac_states}(b)(ii). This is because a switch from $\delta_t(z) \rightarrow -\delta_t(z)$ is equivalent to a switch about the centre and an offset of $\delta_{t,max}$ being added. In a three level system this frequency shift can be overcome by altering the coupling field frequency in such a way as to cancel the initial shift. It should also be noted that $\bar{\delta}_{1,2}$ will not contribute to $\bar{\delta}_t$ for this switching method as it only depends on the detuning and intensity of the acS laser, both of which will be constant.\\
This second switching method would suggest itself as the easiest to implement as all that is required to switch $\eta$ would be a Pockels cell, which have switching times down to nanoseconds. It would not, however, allow for different frequency gradients and therefore filtering or manipulation of the pulse. Apart from this, if the second switching method is used, then the minima for the scattering rate found in Section \ref{sec:ac_wavelength} will only apply for either the read or write stages of the memory. This is because, if the polarisation is reversed, the scattering behaviour of the levels will be reversed (i.e. $\bar{\Gamma}_{2,1}(q=-1) \rightarrow \bar{\Gamma}_{2,-1}(q=1)$.\\
The first switching method would allow for different gradients but involves much longer switching times (on the order of milliseconds for LCSLMs). The combination of beam shapers (BSh) and Pockels cells (PC) shown in Figure \ref{fig:ac_setup}(a) allow for flexibility in beam shaping and fast switching times. If no spatial filtering is desired then only PC2 and BSh1 are needed, with the beam shaper determining the shape of the gradient and the polarisation switch causing the rephasing of the atoms. To allow different gradients to be used, an extra Pockels cell PC1 and beam shaper BSh2 can be used. In this case, the second gradient can be prepared in advance and PC1 used to select which beam shaper to use. The acousto-optic modulator (AOM) can be used to switch the acS beam on or off to decrease the scattering rate due to this field (see following section).\\
\section{Limiting Factors}
\label{sec:ac_limiting_factors}
\subsection{Time Scales}
\label{sec:ac_timescales}
One main advantage of moving from warm to cold atoms is the extended storage times that can be achieved. There are two timescale of importance to our memory: (i) the trap lifetime $\tau_{trap}$; and (ii) the coherence time $\tau_{coh}$.\\
The trap lifetime depends on both the scattering rate of the trapping laser, as well as the rate of inter-atomic collisions. These will also affect the coherence time. Using the trap parameters from Section \ref{sec:ac_atom_trapping} gives a coherence time of approximately 1/$\Gamma_t = 250$ ms. Coupling the trap scattering rate with the recoil energy per emission $E_{rec} = (\hbar k)^2/2m$ \cite{rb11}, where $m$ is the mass of rubidium and $k$ is the wavevector of the transition (here taken to be the D2 transition), give the trap lifetime for a given depth $U_t$ to be
\begin{equation}
\tau_{trap} = \frac{m U_{t}}{\hbar^2k^2\Gamma_{t}}.
\label{eq:ac_trap_lifetime}
\end{equation}
Using this equation and trap parameters determined previously (see Section \ref{sec:ac_GEM_overview}) we find that the trap lifetime will be on the order of 10 seconds and should therefore not affect the coherence time of the system (see below).\\
Another effect that must be considered is the inelastic collision rate between the atoms in the trap, as this will also cause a loss of coherence as well as trap population if the atoms are not all situated in the $F=1$ level. The collision rate in cold atoms has been studied in depth (see for example \cite{mot27, mot26, mot6, mot38}) with the collisional loss rate of the trap being given by
\begin{equation}
\frac{dN}{dt} = -\alpha N - \beta \int n^2(\mathbf{r},t)d^3r,
\label{eq:ac_trap_loss_rate}
\end{equation}
where $n(\mathbf{r},t)$ is the density profile of atoms. As can be seen from the above equation there are two components to the decay: one due to background gas collisions with coefficient $\alpha$ being determined by the background gas pressure; and one due to inelastic collisions between atoms in the trap with coefficient $\beta$. $1/\alpha$ for a dipole trap can, in general, be approximated to 1 s for a trap pressure of $3 \times 10^{-9}$ mbar \cite{rb11}.\\
$\beta$ is perhaps more interesting as it can give an indication of the time between inelastic collisions of the Rb atoms themselves. By using low trap laser intensities, the rate of hyperfine changing collisions has been estimated to be between $\beta_{hcc} = 10^{-11} \rightarrow 10^{-10}$ cm$^3$s$^{-1}$ \cite{mot27,mot38}. $\beta_{hcc}$ will be a lower bound on the total collision rate (as there are also non-hyperfine changing collisions to take into account) but will be used to give an approximate rate of collisions. Here we will take $\beta_{hcc} = 5 \times 10^{-11}$ cm$^3$s$^{-1}$. For the densities we are expecting ($n \approx 10^{11}$ atoms/cm$^3$), and assuming a constant density within the trapping volume for simplicity, Equation \ref{eq:ac_trap_loss_rate} will give an initial, and maximum, collision rate of $\beta n \simeq 30$ s$^{-1}$.\\
Apart from the two sources of decoherence mentioned above, which will also affect the trap lifetime, there are two others that must be considered, namely the scattering rate due to the acS field as well as the coupling field. The scattering rate for the acS field can be determined as a function of system bandwidth $\mathcal{B}_s$ similarly to Equation \ref{eq:ac_optimal_scatering} to be
\begin{equation}
\Gamma_{ac}(\Delta) = \frac{\bar{\Gamma}_{1,-1}(\Delta,1)}{\bar{\delta}_t(\Delta)} \mathcal{B}_s.
\label{eq:ac_scattering_rate_bandwidth}
\end{equation}
For the optimal detuning of $\Delta_{1/2,2}/2\pi = 5$ THz discovered in Section \ref{sec:ac_wavelength} for the $F=1,m_F=-1$ state with $q=1$, this simplifies to $\Gamma_{ac} \simeq 7 \times 10^{-6} \mathcal{B}_s$ for $\bar{\delta}_t \simeq 3 \bar{\delta}_F$.\\
The optical depth of the system depends on the Rabi frequency of the coupling field $\Omega_c$ and the one-photon detuning $\Delta_{1p}$, as was shown in Equation \ref{eq:ac_optical_depth}. It therefore makes sense to express the effect of the coupling field on the ground state in terms of these two parameters by using the relation between coupling field intensity $I_c$ and Rabi frequency $I_c=2 \hbar^2 \epsilon_0c \Omega_c^2/\mu_{23}^2$ to give
\begin{equation}
\Gamma_{c}(\Delta_{1p},\Omega_c)=\frac{2 \hbar^2 \epsilon_0 c}{\mu_{23}^2} \bar{\Gamma}_{1,-1}\left(\Delta_{1p}-\Delta_{hfs},q_c\right) \Omega_c^2.
\label{eq:ac_control_scattering_rate}
\end{equation}
For pulses much longer than 1 $\mu$s and $\left|\Omega_c/\Delta_{1p}\right| \geq 0.001$, then $\Gamma_c \gg \Gamma_{ac}$.\\
The total scattering rate for the system is simply the sum of the individual rates. The acS and coupling fields are only needed during the reading and writing phases of the memory process which each last for a minimum period determined by the pulse length, i.e. $t_{r/w} \geq t_p$, for single pulses but will become longer for multiple pulse storage. If we define a background scattering rate $\Gamma_{bg}$ to include all the decoherence effects that are constantly present, i.e. scattering from the trapping laser, collisions and loss from the trap, and a read/write scattering rate of $\Gamma_{rw} = \Gamma_{ac} + \Gamma_c$, then we can determine the total storage efficiency to be
\begin{equation}
\epsilon_{s}(t_p,\tau) \leq exp\left(-2 t_p \Gamma_{rw}\right) exp\left(-(2t_p +t_s) \Gamma_{bg}\right),
\label{eq:ac_coherence_efficiency}
\end{equation}
where $t_s$ is the time the pulse is stored in the memory.
\subsection{Efficiency}
\label{sec:ac_efficiency}

\begin{figure}[!ht]
\begin{center}
\includegraphics[width=\columnwidth]
{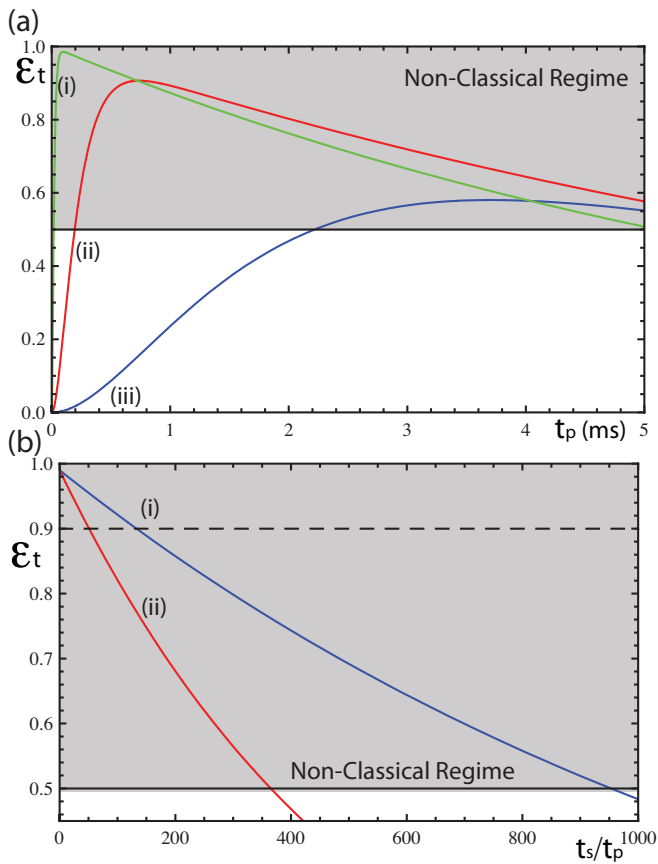} 
\caption{Memory Efficiency. (a) Total efficiency of the $\Lambda$-GEM memory as a function of pulse length $t_p$ for storage times of $t_s = t_p$ and different ratios of $|\Omega_c/\Delta_{1p}|$ = (i) 0.01, (ii) 0.003, and (iii) 0.001. (b) Total efficiency for storage of pulses of length $t_p = 20$ $\mu$s as a function of normalised storage time $t_s/t_p$ for (i) single pulse; and (ii) multiple pulse storage, with $|\Omega_c/\Delta_{1p}| = 0.02$. For all traces $\Delta_{1p} = -2\pi \times 2$ GHz, $\Delta_{ac} = -2\pi \times 5$ THz, $\mathcal{B}_s = \mathcal{B}_G = 9\sqrt{2}/(\pi t_p)$, $q_c=1$, $q_p = 0$, $g = 2 \pi \times 1.5$ MHz, $N = 2.5 \times 10^6$ atoms, $\epsilon_L = 0.4$, $L=1$ cm and $R=10$ $\mu$m.}
\label{fig:ac_efficiency}
\end{center}
\end{figure}

There are three main experimental factors that will affect the storage efficiency of this system: the number of atoms $N$ initially present in the MOT; the loading efficiency from the MOT to the dipole trap $\epsilon_L$; and the size of the gradient which is applied $\left|\eta(z)\right|=2\pi \mathcal{B}_s/L$. Substituting these values into Equations \ref{eq:ac_optical_depth} and \ref{eq:ac_readwrite_efficiency} gives the read/write efficiency of the system
\begin{equation}
\epsilon_{rw} = \left[1-exp\left(-\frac{g^2 \epsilon_L N L}{c \mathcal{B}_s}\left(\frac{\Omega_c}{\Delta_{pc}}\right)^2\right)\right]^2.
\label{eq:ac_total_efficiency}
\end{equation}
To investigate the maximum efficiency as a function of storage time we assume, to maximize optical depth and minimize scattering, that the bandwidth of the system is equal to the bandwidth of the pulse, i.e. $\mathcal{B}_s = \mathcal{B}_p$. For Gaussian pulses we can define the pulse bandwidth to be $\mathcal{B}_G = 9\sqrt{2}/(\pi t_p)$, assuming 99\% of the electric field is stored in the memory. Figure \ref{fig:ac_efficiency}(a) shows the total efficiency $\epsilon_t = \epsilon_{rw} \epsilon_{s}$ for different values of $\Omega_c/\Delta_{1p}$, with storage time $t_s = t_p$, and using $\mathcal{B}_s = \mathcal{B}_G$.\\
As can be seen from the figure, for each ratio of $\Omega_c/\Delta_{1p}$ and storage time there is a maximum efficiency. This is a combination of two effects: firstly the increase in optical depth as the pulse length increases due to the smaller bandwidths required, and extra decoherence that will occur as the storage time increases. The decoherence is greater for larger values of $\Omega_c/\Delta_{1p}$ due to the increase in $\Gamma_{rw}$ this entails, and this can also be seen from Figure \ref{fig:ac_efficiency}(a). In this regime, however, the main source of decoherence is the atomic collisions, with an approximate coherence time of 30 ms.\\ 
For short pulses ($t_p \ll 1$ ms) and large values of $\left|\Omega_c/\Delta_{1p}\right| \simeq 0.02$ efficiencies approach unity as optical depths will be high and the background decoherence effects will be negligible. This is therefore a good regime to investigate the delay-bandwidth product (DBP) of the memory. Here we define DBP$\equiv t_s/t_p$ as this will give an indication of the number of pulses (or bits) that can be stored in the memory at one time.\\
Figure \ref{fig:ac_efficiency}(b) shows $\epsilon_t$ for (i) single and (ii) multiple pulse storage with $t_p = 20$ $\mu$s. The difference between the two is that, for multiple pulse storage, the coupling and acS fields are left on at all times, while this is not the case for single pulse storage. As can be seen, single pulse storage with efficiency greater than 90\% (dashed line) occurs up to approximately $t_s = 130 t_p$. For multiple pulse storage, this drops to 50 pulses due to the extra coupling field scattering, while 350 can be stored above the classical efficiency limit of $\epsilon_t = 0.5$. Currently a maximum of four pulse storage has been achieved with $\Lambda$-GEM with $\epsilon_t \simeq 1$ \cite{g5}. To store pulses of 20 $\mu$s length requires a system bandwidth of approximately 200 kHz and therefore an acS laser power of less than 2 W in the optimized regime.
\section{Conclusions}
We have made two proposals to improve the current experimental implementation of $\Lambda$-GEM. First, a new method of gradient creation using the ac Stark effect. Second, a move from warm to cold atoms, along with the longer coherence times this entails.\\
This paper investigated the experimental viability of these proposals. Firstly, the trapping mechanism was investigated and it was determined that a far-detuned standing-wave dipole trap would not interfere with the acS gradient and have a lifetime on the order of seconds, while providing a small area for the acS laser. It was found that there is an optimal detuning for the acS field of approximately 5 THz to minimize scattering and maximize bandwidth. With this detuning, and an optimized beam shape and level scheme, a bandwidth of 150 kHz/W could be created. Different methods for gradient switching were investigated and a scheme devised that would allow a switch between any two arbitrary gradients on the order of nanoseconds using Pockels cells.\\
Finally, factors that would limit the system such as scattering due to the trapping, coupling and acS lasers, as well as collisions between atoms were examined. These were combined with the read/write efficiency to model the total system efficiency as a function of storage time and it was seen that for long pulses ($t_p \gg 1$ ms), the coherence time was limited by the atomic collisions to be on the order of 10s of milliseconds. It was also found that for short pulses ($t_p \ll 1$ ms), efficiencies approach unity and 50 pulses could be stored in the memory at one time with efficiency greater than 90 \%. The acS laser power required to produce the necessary memory bandwidth to store these pulses would be less than 2 W in the optimized regime.\\
We therefore conclude that using an ac Stark gradient with cold atoms is an experimentally viable option for improving the $\Lambda$-GEM system in terms of gradient creation, switching and manipulation, as well as storage times and time-bandwidth products achievable using cold atoms.

\section*{Acknowledgements}
We thank D. D\"{o}ring for enlightened discussions and his rubidium ac Stark shift code. This work was supported by the Australian Research Council.

\clearpage


\begin{thebibliography}{99}
\bibitem{rwa}
Calculations included the counter-rotating terms ($\omega_0 + \omega_l$) as rotating wave approximation not valid for large detunings where $\omega_o - \omega_l \ll \omega_o + \omega_l$ no longer holds.
\end{thebibliography}
\end{document}